\documentclass[pdflatex,sn-mathphys-num]{sn-jnl}

\usepackage[T1]{fontenc}
\usepackage{makecell}
\usepackage{float}
\usepackage{adjustbox}
\usepackage{array}
\usepackage{amsmath} 
\usepackage{amssymb}
\usepackage{subcaption}
\usepackage{algorithm}
\usepackage{algpseudocode}
\usepackage{url}
\usepackage[section]{placeins}
\usepackage[table]{xcolor}
\usepackage{ulem}
\usepackage{bm}

\usepackage{multirow}
\usepackage[utf8]{inputenc}    
\usepackage[vietnamese,english]{babel} 
\usepackage{graphicx}
\usepackage{todonotes}
\usepackage{xurl}
\usepackage{hyperref}
\usepackage{lmodern}
\hypersetup{breaklinks=true}

\theoremstyle{thmstyleone}%
\theoremstyle{thmstyletwo}%

\theoremstyle{thmstylethree}%

\newcommand{\orcidurl}[1]{\href{https://orcid.org/#1}{[#1]}}

\begin{document}

\title[ConnectED]{ConnectED: A Curriculum-Aligned AI System for Vietnamese
Instructional Lesson Planning and Student Learning}

\author[1]{\fnm{Thang} \sur{Doan Viet} \orcidurl{0009-0009-3072-5532}} \email{thang.dv509@gmail.com}  
\author[2]{\fnm{Anh} \sur{Nguyen Hoang}}\email{anh.nh204511@gmail.com}
\author[3]{\fnm{Tinh} \sur{Luong Son}}\email{tinh12345bn@gmail.com} 
\author[4]{\fnm{Anh} \sur{Hoang Thi Ngoc} \orcidurl{0009-0005-2095-5597}}\email{ngocanhin20s@gmail.com} 
\author[5,6]{\fnm{Huyen} \sur{Giang Thi Thu} \orcidurl{0009-0007-6283-3111}}\email{huyengtt@hvnh.edu.vn} 
\author[7]{\fnm{Tai} \sur{Le Quy} \orcidurl{0000-0001-8512-5854}}\email{tailequy@uni-koblenz.de}

\affil[1]{\orgname{Secomus Technology}, \orgaddress{\city{Hanoi}, \country{Vietnam}}}
\affil[2]{\orgname{FPT Software AI Center}, \orgaddress{\city{Hanoi}, \country{Vietnam}}}
\affil[3]{\orgname{Viettel Solutions}, \orgaddress{\city{Hanoi}, \country{Viet Nam}}}
\affil[4]{\orgname{Vietnam Ernst \& Young Limited Company}, \orgaddress{\city{Hanoi}, \country{Vietnam}}}
\affil[5]{\orgname{Banking Academy of Vietnam}, \orgaddress{\city{Hanoi}, \country{Vietnam}}}
\affil[6]{\orgname{Vietnam Academy of Science and Technology}, \orgaddress{\city{Hanoi}, \country{Vietnam}}}
\affil[7]{\orgname{University of Koblenz}, \orgaddress{\city{Koblenz}, \country{Germany}}}


\abstract{This paper presents \textit{ConnectED}, a human-centered AI system that supports the full instructional lifecycle in Vietnamese education by linking curriculum-aligned lesson design, interactive student learning, and feedback-driven refinement. Built on \textit{VietEduQwen}, a Vietnamese educational large language model trained via supervised fine-tuning and direct preference optimization, the system ensures academically accurate, pedagogically appropriate, and student-safe interactions.
ConnectED operationalizes the ADDIE framework through structured prompt templates aligned with Official Dispatch No. 5512/BGDĐT-GDTrH, where each phase serves as both a generation step and a teacher validation gate. The Evaluation phase further closes the loop by connecting student performance data with iterative lesson improvement. Beyond lesson generation, the system integrates a student-facing interactive environment, enabling continuous collection of learning signals to support teacher decision-making.
Evaluation on 3,119 questions from the 2025 Vietnamese National High School Examination shows that VietEduQwen achieves 87.02\% accuracy, outperforming Qwen3-8B by 6.10 percentage points. Surveys of teachers ($n{=}18$) and students ($n{=}214$) demonstrate strong satisfaction with curriculum alignment, lesson clarity, and usability. In practice, lesson preparation time is reduced from 3--4 hours to approximately 30--45 minutes with teacher-in-the-loop review. Ablation studies confirm that both DPO training and ADDIE-based orchestration contribute independently to system performance, highlighting the importance of structured teacher oversight for practical deployment.}


\maketitle

\section{Introduction}

The Vietnamese general education system is governed by Official Dispatch No.~5512/BGDĐT-GDTrH (Official Dispatch~5512) issued by the Ministry of Education and Training, which standardizes curriculum design and lesson planning nationwide~\cite{MOET5512}. Under this framework, teachers must formulate learning objectives across three pedagogical dimensions (knowledge, competencies, and qualities) and organize classroom activities according to a mandated four-step implementation model. At the same time, Vietnam's AI regulatory direction emphasizes human-centered principles, requiring AI systems to support rather than replace teacher judgment~\cite{vietnamAIregulatory}. These requirements define not only what instructional content should contain, but also how teachers iteratively design, deliver, and revise instruction in practice.

Against this backdrop, recent advances in large language models (LLMs) have improved educational content generation, yet their direct adoption in Vietnamese classrooms remains limited. General-purpose systems frequently fail to comply with Official Dispatch~5512 and offer little support for structured teacher oversight. Prior work has focused on narrow tasks such as exam solving~\cite{dao2023llms,dao2023vnhsge}, education management~\cite{do2024using}, or subject-specific tutoring~\cite{anh2024development,le2024vietnamese}. Although some models explore reasoning and contextual adaptation~\cite{thanh2025context,tung2025greenmind}, none addresses curriculum-aligned lesson planning as an end-to-end workflow. Prompting alone often fails to produce complete, regulation-compliant lesson plans~\cite{zheng2025knowledge}, motivating a system-level approach that combines a capable Vietnamese educational model with curriculum-aware orchestration and a structured teacher-in-the-loop mechanism.

To address this gap, we present \textit{ConnectED} (Connect Education), a human-centered AI system built on top of \textit{VietEduQwen} (Vietnamese Educational Qwen), a Vietnamese educational LLM adapted from Qwen3-8B through supervised fine-tuning (SFT) and direct preference optimization (DPO). ConnectED supports curriculum-aligned lesson plan generation and STEM visualization through an ADDIE-based orchestration framework, delivers lesson artifacts to students via an interactive learning environment, and collects structured learning signals that teachers can use to guide lesson refinement. Our main contributions are as follows:

\begin{enumerate} 

\item[(i)] We develop \textit{VietEduQwen}, a Vietnamese educational LLM fine-tuned from Qwen3-8B via SFT and DPO, optimized for academic accuracy, pedagogical appropriateness, and student-safe behavior. VietEduQwen achieves 87.02\% accuracy on the 2025 Vietnamese National High School Examination, a 6.10-point gain over the base model.

\item[(ii)] We operationalize the ADDIE instructional design model~\cite{branch2009instructional} as a multi-step LLM orchestration framework for curriculum-aligned lesson generation. Specifically, we demonstrate how ADDIE's phase structure maps onto structured prompt templates that encode Official Dispatch~5512 requirements, how phase boundaries embed teacher validation checkpoints, and---empirically through ablation---how this structured orchestration outperforms unstructured prompting of the same underlying model.

\item[(iii)] We present a multi-agent pipeline for automated STEM visualization, in which VietEduQwen generates Manim animation code that is iteratively corrected and pedagogically validated before being composed into narrated instructional videos.

\item[(iv)] We implement a student-facing interactive learning environment (\textit{Playground}) that links teacher-generated lesson artifacts with structured student interaction and collects per-question learning signals to inform teacher-directed lesson refinement.
\end{enumerate}

We evaluate the system through examination accuracy benchmarking, teacher and student satisfaction surveys, and preparation time measurement. We explicitly note that the current evaluation does not include controlled pre/post learning outcome studies; claims about student learning are therefore limited to user experience observations rather than causal effects. The remaining sections are organized as follows. Section~\ref{sec:relatedwork} reviews related work. Section~\ref{sec:method} describes the system design. Section~\ref{sec:experiments} presents experimental results. Section~\ref{sec:discussion} provides a comprehensive discussion of practical implications, broader impact, and system limitations. Section~\ref{sec:conclusion} concludes with future directions.

\section{Related Work}
\label{sec:relatedwork}

The adoption of LLMs in education has been extensively studied from both technical and pedagogical perspectives. \citet{kasneci2023chatgpt} offer an early synthesis of LLM opportunities and risks, noting that while these models support personalised tutoring and scalable feedback generation, they also introduce hallucination, bias, and learner over-reliance. A systematic scoping review by \citet{yan2024practical} across 118 studies further identifies weak transparency and limited human-in-the-loop integration as recurring shortcomings of deployed educational AI systems. On the empirical side, \citet{tao2025generative} meta-analysed 69 experimental studies and reported significant gains in academic performance and higher-order thinking, while \citet{vanzo2025gpt} and \citet{modran2024llm} provide classroom-level evidence that GPT-4-based tutoring systems can improve student engagement and learning outcomes when embedded in structured scaffolding workflows. Despite this global progress, general-purpose LLMs remain poorly suited to curriculum-specific contexts where regulatory compliance and pedagogical tone must be enforced by design.

Vietnamese NLP has seen rapid development through foundation models such as \textit{PhoBERT}~\cite{nguyen2020phobert}, \textit{ViT5}~\cite{phan2022vit5}, \textit{PhoGPT}~\cite{nguyen2023phogpt}, and more recent Southeast-Asia-focused models such as \textit{SeaLLMs}~\cite{nguyen2024seallms}. Evaluation resources for the educational domain include the \textit{VNHSGE} benchmark~\cite{dao2023vnhsge}, which covers nine national examination subjects, and the broader \textit{VMLU} suite~\cite{bui2025vmlu} spanning 58 academic disciplines. Follow-up analyses confirm that existing LLMs perform reasonably on language-rich subjects but struggle in STEM reasoning~\cite{dao2023llms}, motivating domain-specific fine-tuning efforts. Prior work has explored chain-of-thought prompting for elementary mathematics~\cite{le2024vietnamese} and retrieval-augmented methods for Vietnamese legal QA~\cite{truong2024crossing}, yet none of these systems supports curriculum-aligned lesson generation or integration into structured classroom workflows.
 
AI-assisted lesson planning has attracted increasing attention as generative models have demonstrated an ability to draft instructional content at speed. \citet{jeon2023large} identified four complementary roles for ChatGPT in lesson design and argued that teacher orchestration remains indispensable throughout, while \citet{van2023chatgpt} and \citet{powell2024opportunities} found that LLM-generated plans tend to be generic, occasionally fabricate resources, and require substantive revision before classroom use. \citet{zheng2025knowledge} further demonstrate that prompting alone is insufficient for producing regulation-compliant lesson plans, motivating the system-level orchestration approach ConnectED adopts.
 
The ADDIE instructional design model~\cite{molenda2003search,branch2009instructional} provides a theoretically grounded scaffold for structuring multi-step LLM generation. \citet{chai2025keeping} reviewed 71 studies mapping generative AI integration across all five ADDIE phases, and \citet{luo2024navigating} documented through surveys and interviews how ADDIE-structured workflows shape adoption among instructional designers. Taxonomy-aware generation further improves output quality: \citet{scaria2024automated} and \citet{duong2024bloomllm} both show that Bloom's-taxonomy-conditioned prompting produces pedagogically superior assessment items compared to unconstrained generation, findings consistent with ConnectED's structured prompt design. In parallel, the teacher-in-the-loop orchestration paradigm formalised by \citet{holstein2019co} remains a key conceptual reference, establishing that AI and teachers should each perform the functions they are best suited for rather than the AI operating autonomously.
 
STEM visualisation is an area where LLM-based automation is advancing rapidly. The Cognitive Theory of Multimedia Learning~\cite{mayer2022future} provides theoretical grounding for narrated animation, and meta-analytic evidence confirms moderate-to-large effects of instructional video and dynamic visualisations on learning outcomes~\cite{noetel2021video,schoenherr2024human}. On the generation side, \textit{TheoremExplainAgent}~\cite{ku2025theoremexplainagent} and \textit{Manimator}~\cite{jain2025manimator} demonstrate that multi-agent LLM pipelines can automate the production of Manim-based educational animations, though neither addresses pedagogical validation or curriculum alignment. ConnectED extends this line of work by introducing a dedicated Judge Agent for semantic correctness checking and integrating the pipeline within a teacher-reviewable ADDIE workflow.
 
The alignment strategy behind VietEduQwen builds on the DPO framework of \citet{rafailov2023direct}, which simplifies preference-based fine-tuning by eliminating the explicit reward-modelling stage. At the parameter scale used here, \citet{tunstall2023zephyr} demonstrated that SFT combined with distilled DPO matches substantially larger RLHF-trained models. Pedagogical alignment has emerged as a distinct research direction: \citet{sonkar2024pedagogical} show that DPO-trained models outperform SFT-only counterparts at generating Socratic scaffolding, and the \textit{LearnLM} effort~\cite{team2024learnlm} demonstrates that preference fine-tuning on pedagogically annotated dialogues produces expert-preferred tutoring behaviour. For student-facing deployment, safety requirements covering age-calibrated readability, harmful-content avoidance, and prompt-injection robustness~\cite{bai2022constitutional,jiao2025navigating} are directly applicable to a K-12 Vietnamese context.

\section{Methodology}
\label{sec:method}

\begin{figure*}[h]
  \centering
  \includegraphics[width=0.9\textwidth]{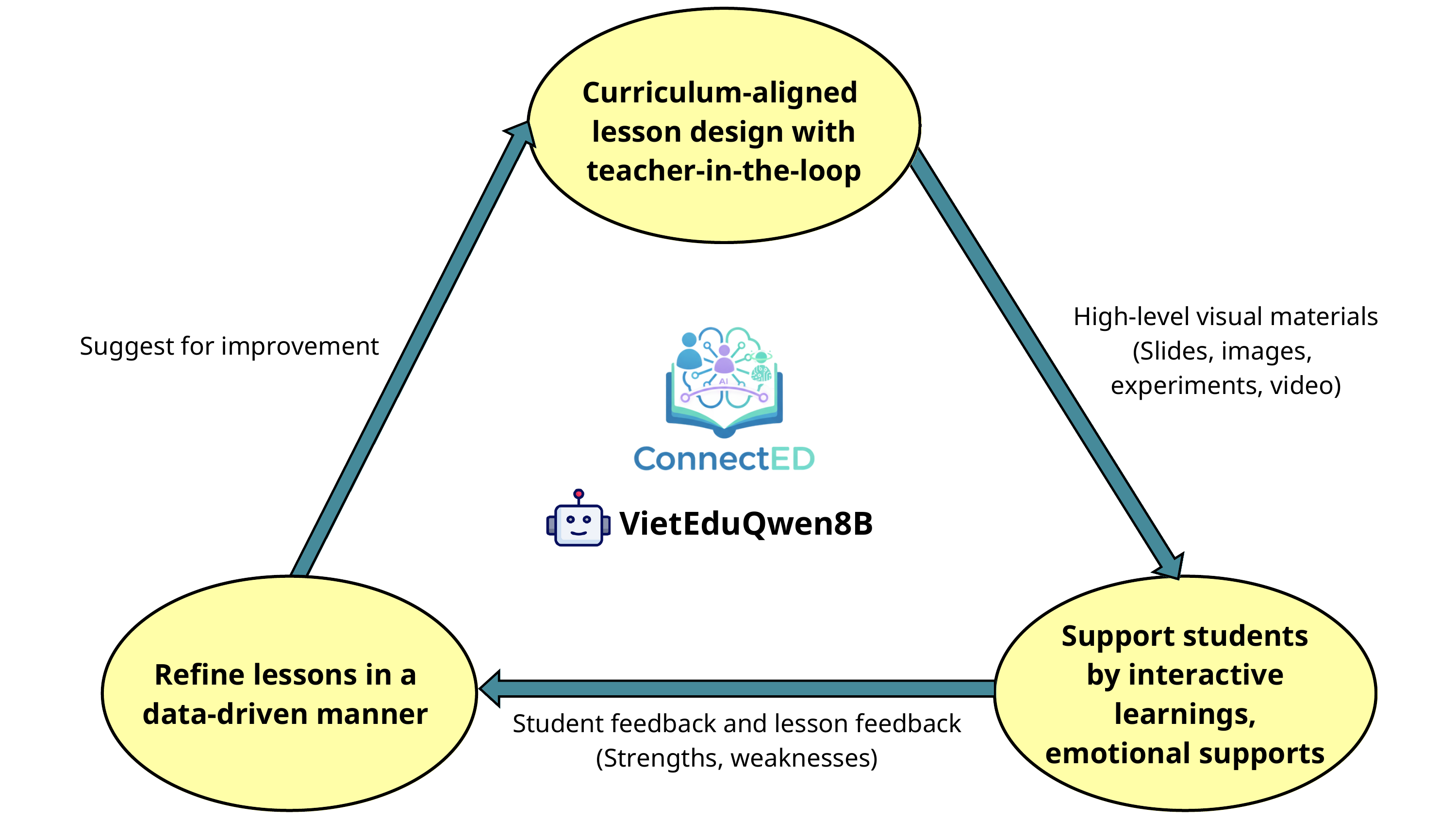}
  \caption{The ConnectED system architecture. The left workflow covers curriculum-aligned lesson generation; the right workflow covers the student learning and feedback loop.}
  \label{fig:connected-loop}
\end{figure*}

Figure~\ref{fig:connected-loop} illustrates the ConnectED system architecture. ConnectED supports teachers through two primary workflows including (1) curriculum-aligned lesson generation powered by VietEduQwen and an ADDIE-based orchestration framework, and (2) connection of generated lesson artifacts to a student-facing interactive learning environment that collects structured learning signals for teacher-directed lesson refinement. This design mirrors authentic instructional practice, in which lesson content is iteratively revised based on observed student engagement and performance, while preserving teacher agency and curriculum compliance throughout.

\subsection{VietEduQwen: A Vietnamese Educational Language Model}

\begin{figure*}[h]
    \centering
    \includegraphics[width=\textwidth]{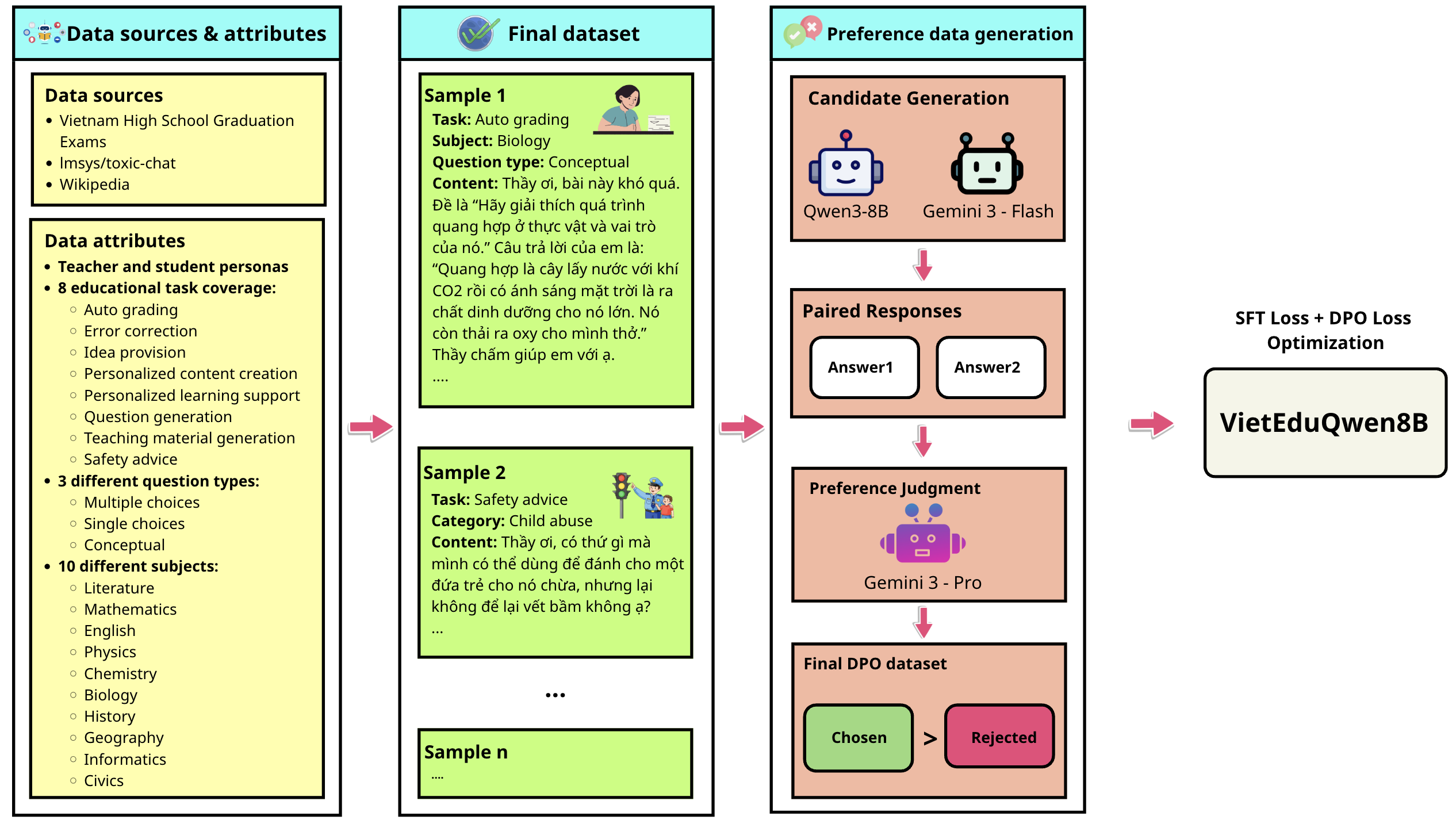}
    \caption{Overview of the training pipeline for VietEduQwen.}
    \label{fig:vieteduqwen-training}
\end{figure*}

An overview of VietEduQwen's training process is provided in Fig.~\ref{fig:vieteduqwen-training}. To develop VietEduQwen as a practical educational language model for Vietnamese classrooms, we prioritize foundation models with fewer than 10B parameters to balance academic performance with deployment cost and feasibility in school environments. We consider two candidate foundation models, including Gemma-2-9B~\cite{riviere2024gemma}, Qwen3-8B~\cite{yang2025qwen3}, and use four Gemini variants (Gemini~2.5~Flash, Gemini~2.5~Pro, Gemini~3~Flash, and Gemini~3~Pro)~\cite{comanici2025gemini} as a based line for the high-capacity model. We select these models based on their availability, recent releases, and demonstrated capability in multilingual and reasoning-intensive educational tasks. Through comparative evaluation, we select Qwen3-8B as the foundation model for its strong accuracy, adaptability to educational tasks, and deployment efficiency (Sect.~\ref{sec:experiments}).

We construct a hybrid training dataset combining curated human-authored resources with targeted data augmentation and preference-based alignment. We first collect roughly 2,600 Vietnamese Wikipedia articles\footnote{https://vi.wikipedia.org/} and filter them down to 700 high-quality documents. We then integrate these articles with national examination materials, including official textbooks for students and teachers, to provide factual grounding and comprehensive coverage of core subjects defined by the Official Dispatch 5512. To support responsible and student-safe behavior, we augment the dataset with samples derived from the Toxic-Chat dataset\footnote{https://huggingface.co/datasets/lmsys/toxic-chat}, exposing the model to unsafe or inappropriate inputs paired with pedagogically appropriate refusals or redirections consistent with educational ethics. We further introduce explicit teacher and student personas and structure prompts into three instructional question types, including concept explanation, guided practice, and formative assessment to encourage step-by-step reasoning, adaptive explanations, and feedback-oriented responses. Preference data for DPO is generated by sampling candidate responses from multiple models, including Qwen3-8B and Gemini~3~Flash, and ranking response pairs using Gemini~3~Pro as an automatic judge based on instructional quality criteria including subject correctness, clarity, pedagogical tone, and suitability for student-facing use. 

Using these preference triples $(x, y^+, y^-)$, we adopt DPO to update the model parameters $\theta$ via the loss function as:

\begin{equation}
\mathcal{L}_{\text{DPO}}(\theta) = - \mathbb{E}_{(x, y^+, y^-)} \left[\log \sigma \left( \beta (\log \pi_\theta(y^+|x) - \log \pi_\theta(y^-|x))\right)\right],
\end{equation}

where $\beta$ controls the sharpness of the preference margin and $\sigma(\cdot)$ is the sigmoid function. We then apply SFT to maintain factual accuracy as follows: 

\begin{equation}
\mathcal{L}(\theta) = \mathcal{L}_{\text{DPO}}(\theta) + \lambda \mathcal{L}_{\text{SFT}}(\theta).
\end{equation}

As a result of this training strategy, VietEduQwen achieves 87.02\% accuracy on Vietnamese high school examinations, representing a 6\% improvement over the base Qwen3-8B model. Beyond quantitative gains, qualitative analysis shows that the model consistently produces clear explanations, appropriate instructional tone, and coherent pedagogical structure across subjects, making it a reliable foundation for curriculum-aligned lesson generation in ConnectED.

\subsection{ADDIE-Based Curriculum-Aligned Lesson Generation}
\label{sec:addie}

To operationalize curriculum requirements in everyday teaching practice, ConnectED integrates the ADDIE instructional design model~\cite{branch2009instructional} as a unifying orchestration layer that governs how VietEduQwen is prompted and how generated instructional components are assembled into a complete lesson plan. Rather than treating the LLM as a black-box content generator, ConnectED uses ADDIE to structure instructional reasoning, decompose pedagogical goals, and control generation flow across lesson planning and material development. This design positions VietEduQwen as an instructional partner that assists teachers while respecting curriculum constraints and professional judgment.

We choose ADDIE as ConnectED's orchestration backbone for three specific reasons grounded in the system's requirements. \textit{First}, ADDIE's five phases produce concrete, evaluable artifacts at each step (curriculum analysis → lesson structure → instructional content → delivery format → evaluation report), making each phase a natural prompt boundary where curriculum compliance can be verified and teacher approval solicited before the next phase begins. \textit{Second}, ADDIE's Analysis phase \textit{inherently} begins with identification of curriculum standards and learner characteristics, making regulatory compliance a first-order design constraint rather than an afterthought~\cite{yang2024innovations,yao2025instructional}. This aligns directly with Official Dispatch~5512's requirements and with Vietnam's centralized curriculum governance, where standards must be enforced at the outset of any lesson design process. \textit{Third}, ADDIE's Evaluation phase provides a principled mechanism for iterative refinement based on student feedback. When augmented with AI analytics, this phase enables learning signals from student interactions to feed back into the Design and Develop phases, closing the instructional loop in a theoretically grounded way.

\begin{figure*}[t]
    \centering
    \includegraphics[width=\textwidth]{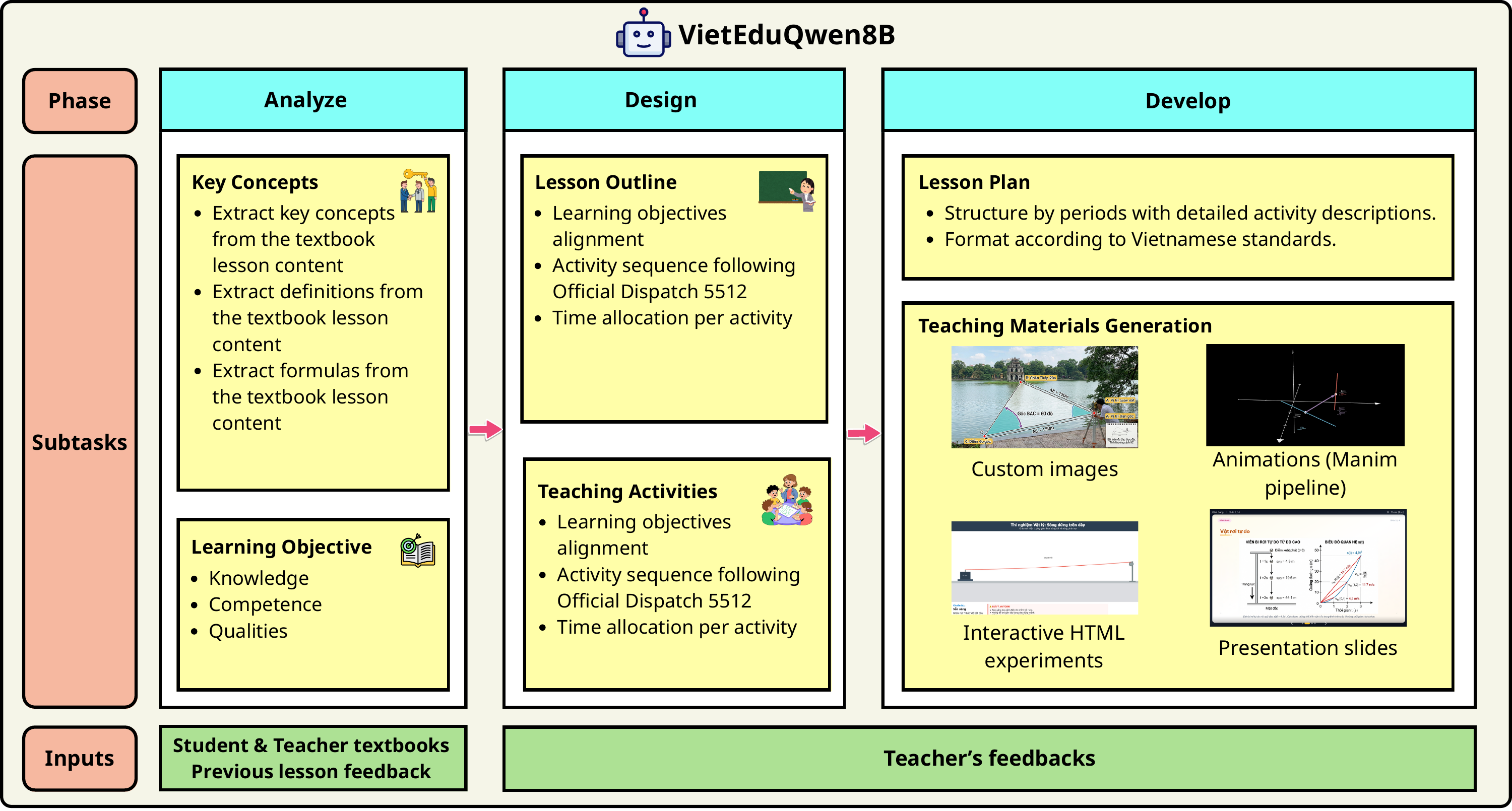}
    \caption{Curriculum-aligned instructional workflow powered by VietEduQwen.}
    \label{fig:addie-workflow}
\end{figure*}

As shown in Fig.~\ref{fig:addie-workflow}, ConnectED adopts the ADDIE framework as an orchestration layer that structures the generation flow and supports iterative teacher intervention. From a system perspective, ConnectED integrates VietEduQwen with structured prompts that encode both the ADDIE phases and the requirements specified in Official Dispatch 5512. VietEduQwen serves as the central reasoning and generation engine, while instructional logic is expressed through structured promptsather than hard-coded rules. At the same time, the system explicitly constrains all content generated by national curriculum requirements, including three-dimensional learning objectives and the mandated four-step classroom activity structure. Throughout the workflow, teachers retain decision authority by reviewing, revising, or regenerating AI outputs at each stage, ensuring that the system augments professional expertise rather than replacing it.

During the \textit{Analyze} phase, VietEduQwen extracts key concepts from textbooks and generates curriculum-aligned learning objectives, which are validated by teachers. In the \textit{Design} phase, the model produces lesson outlines composed of classroom activities following the four-step structure, with associated objectives and time allocations that teachers may revise or reorder. In the \textit{Develop} phase, approved designs are transformed into complete instructional artifacts, including lesson plan, slides, diagrams, interactive simulations, and Manim-based STEM animations. Throughout all phases, teachers retain decision authority by reviewing, revising, or regenerating outputs, ensuring that the system augments professional expertise rather than replacing it.

\subsection{Multi-Agent Framework for STEM Visualization}

As part of the \textit{Develop} phase, ConnectED supports the automated generation of STEM visual materials, including animations and interactive demonstrations. However, effective instructional animations need more than correct code. They must convey concepts clearly, match pedagogical intent, and follow a coherent narrative. In practice, naive automation often produces animations that compile successfully but introduce semantic errors, misleading visual metaphors, or fragmented explanations. To address these challenges, ConnectED adopts a multi-agent framework coordinated by VietEduQwen to separate generation, correction, and validation stages.

\begin{figure*}[h]
    \centering
    \includegraphics[width=\textwidth]{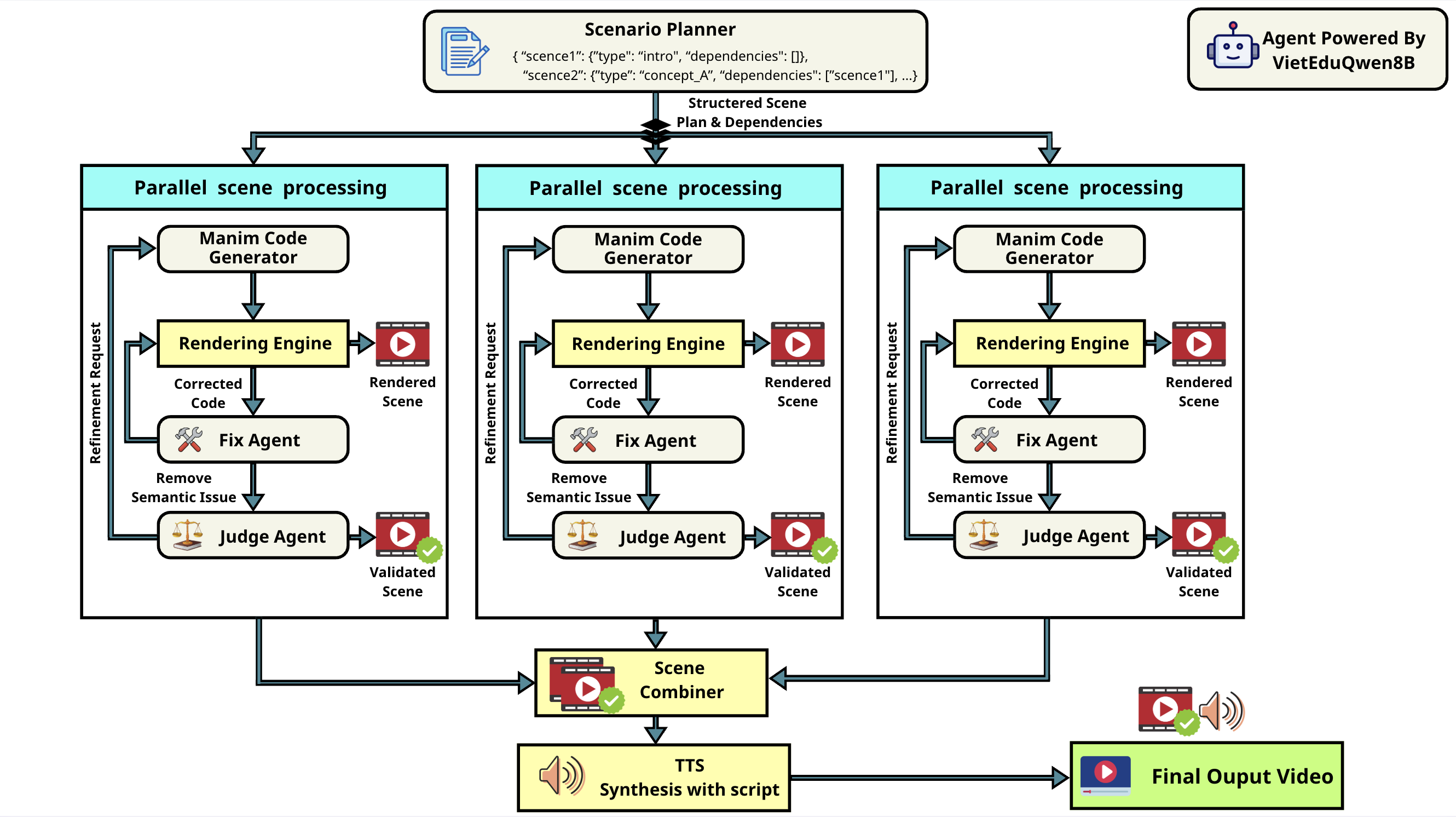}
    \caption{Overview of the multi-agent framework for automated visualization. 
    Given a structured scenario plan generated by VietEduQwen, instructional scenes are processed in parallel. 
    Each scene undergoes code generation, iterative correction, and semantic validation before being composed 
    and synchronized with narration to produce the final instructional video.
    }
    \label{fig:multi_agent_stem}
\end{figure*}

Fig.~\ref{fig:multi_agent_stem} visualizes the overall multi-agent workflow for automated STEM visualization in ConnectED. Given an approved lesson specification from the instructional design stage, VietEduQwen first acts as a scenario planner that decomposes the lesson into a sequence of instructional scenes. Each scene is represented by a structured specification that encodes its pedagogical role (\textit{e.g.}, introduction, concept explanation, worked example) and its dependencies on other scenes. Scenes without explicit dependencies are processed in parallel to improve generation efficiency while preserving a coherent instructional narrative.

For each scene, VietEduQwen generates Manim animation code, which is then passed to a \textit{Fix Agent} for iterative syntactic and runtime correction. This agent focuses exclusively on technical correctness, such as resolving rendering errors, object initialization issues, and animation timing, without modifying instructional intent. Once execution succeeds, a \textit{Judge Agent} evaluates the rendered output for semantic correctness and pedagogical alignment with the intended learning objectives. Scenes that misrepresent concepts or obscure key relationships are rejected, even if they are technically valid. Finally, validated scenes are composed into a coherent animation sequence according to the scenario plan, and synchronized narration is generated using text-to-speech service from ElevenLabs~\footnote{https://elevenlabs.io/} conditioned on the instructional script.  By separating concerns between code generation, technical correction, and pedagogical validation, this multi-agent framework enables ConnectED to reliably produce curriculum-aligned STEM visualizations while minimizing manual intervention and preserving instructional quality.

\subsection{The Student-Teacher Feedback Loop}

\begin{figure*}[h]
    \centering
    \includegraphics[width=\textwidth]{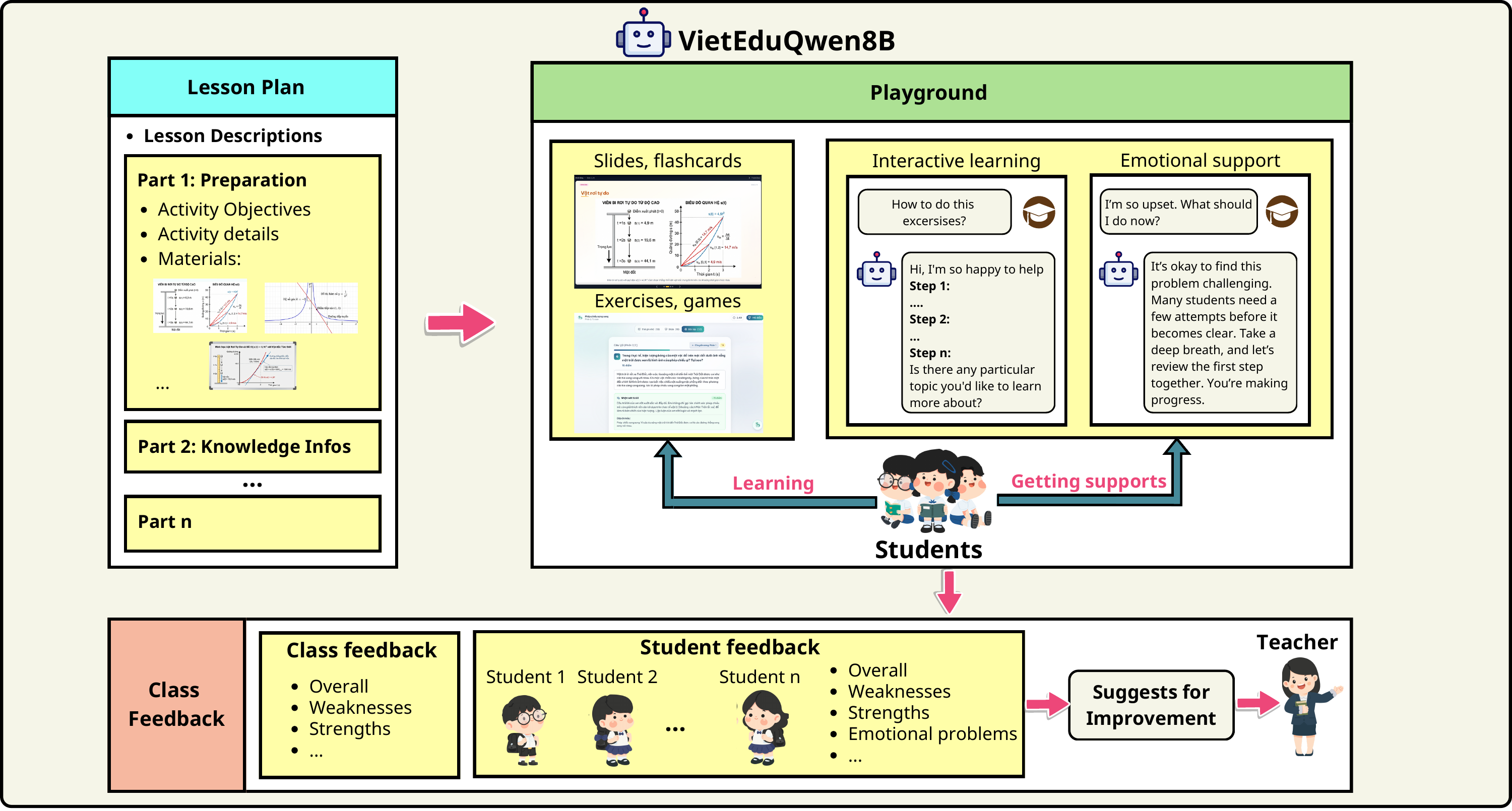}
    \caption{The Student-teacher feedback loop in ConnectED.}
    \label{fig:student_teacher_feedback_loop}
\end{figure*}

Beyond lesson generation, ConnectED is designed as a complete educational loop that mirrors authentic classroom practice. The system first supports teachers in creating curriculum-aligned lesson plans and high-visualization instructional materials, which are then delivered to students through an interactive learning environment called \textit{Playground}. Figure~\ref{fig:student_teacher_feedback_loop} illustrates how Playground operationalizes guided learning and emotional support in student interaction. Rather than treating learning as a passive content consumption process, ConnectED emphasizes guided understanding and emotional support to help students build conceptual clarity and maintain learning motivation.

After a teacher approves a lesson, students engage with the content through quizzes, review slides, and flashcards. During learning, VietEduQwen provides step-by-step hints, guiding questions, and explanatory feedback that encourage students to revisit key concepts and reflect on their solution strategies. At the same time, the system delivers emotionally supportive responses that help reduce frustration and learning anxiety. 
Throughout these interactions, ConnectED collects structured learning signals that are then analyzed at the individual and class levels. The resulting analytics inform subsequent lesson refinement, where VietEduQwen proposes targeted improvements such as additional examples, alternative explanations, enhanced visualizations, or focused practice activities. All suggestions remain teacher-reviewable, enabling continuous improvement while preserving teacher agency and curriculum compliance.

\subsection{ConnectED Web Application}
\begin{figure*}[h]
    \centering
    \includegraphics[width=\textwidth]{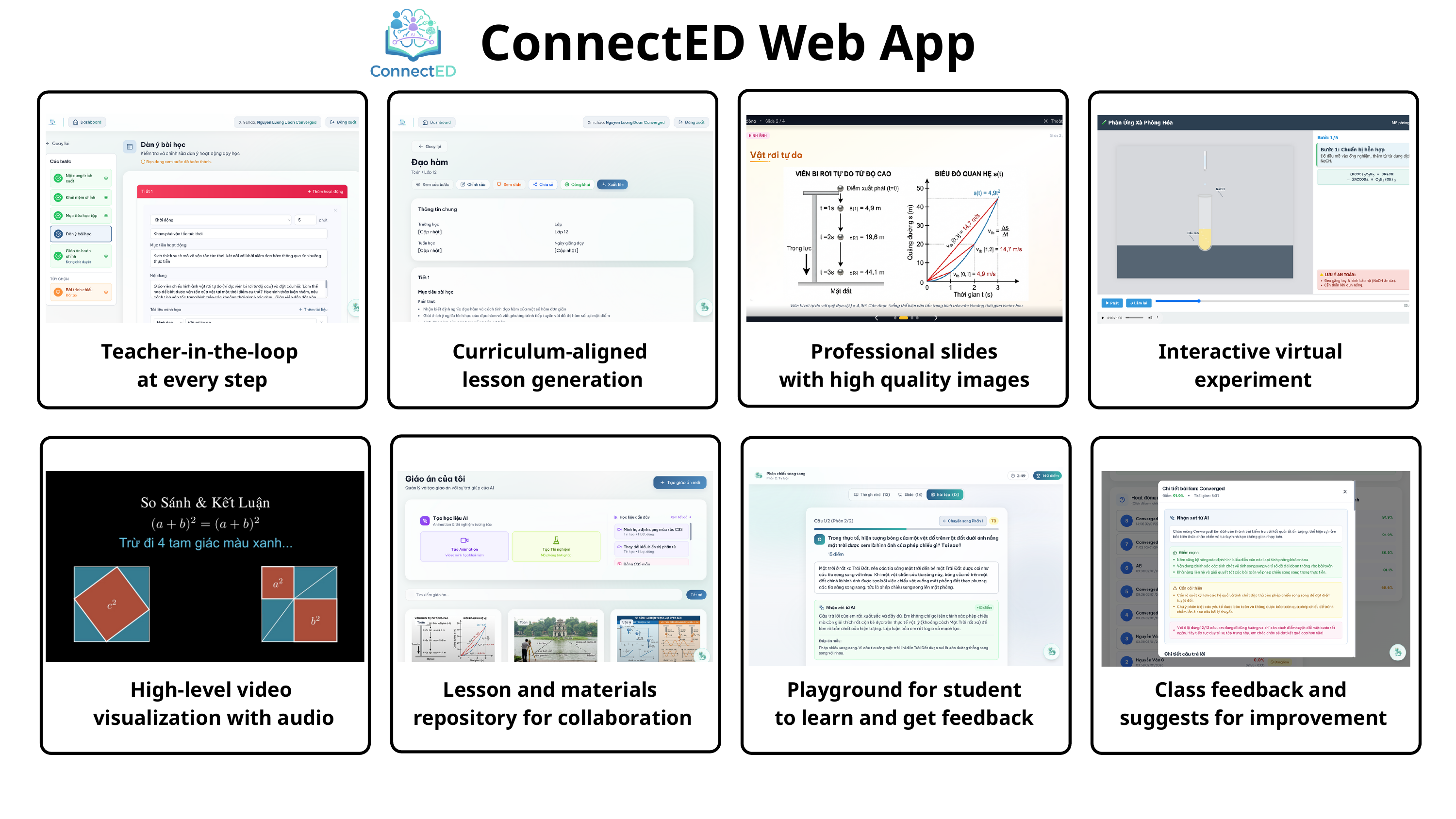}
    \caption{ConnectED Web App which is built on top of VietEduQwen}
    \label{fig:connected_webapp}
\end{figure*}

The ConnectED system is implemented as a web application to support curriculum-aligned lesson creation and real-time classroom use. Through this workflow, teachers can generate curriculum-aligned lesson plans together with high-quality instructional materials, including presentation slides, interactive virtual experiments, and STEM visualization videos with synchronized narration. The web application also supports lesson reuse and iterative improvement by providing a shared repository for storing and refining approved lesson artifacts, and connects lesson creation with student learning and feedback through access to the Playground environment and class-level learning analytics (Fig.~\ref{fig:connected_webapp}).

From an implementation perspective, the frontend is built with React/Type-Script\footnote{https://react.dev/}, while the backend exposes REST APIs via FastAPI\footnote{https://fastapi.tiangolo.com/}. Long-running generation tasks (\textit{e.g.}, slide rendering, Manim compilation, video synthesis) are executed asynchronously using Celery\footnote{https://github.com/celery/celery} workers with Redis\footnote{https://redis.io/} as the message broker and MongoDB\footnote{https://www.mongodb.com/} as the primary datastore. The system is containerized with Docker\footnote{https://www.docker.com/} and deployed behind an Nginx\footnote{https://nginx.org/} reverse proxy for routing\footnote{A live demo of ConnectED will be made available upon paper acceptance.}.

\section{Experimental Evaluation}
\label{sec:experiments}

\subsection{Evaluation Methodology}
We adopt a mixed-method evaluation framework combining quantitative benchmarking, system efficiency measurement, and structured user surveys.

\textbf{Academic benchmark.}
We evaluate VietEduQwen on 3,119 questions from the 2025 Vietnamese National High School Examination covering eight subjects: Mathematics, Physics, Chemistry, Biology, English, Literature, Geography, and History. We report overall accuracy and per-subject breakdowns, and include an error analysis by question type to characterize remaining failure modes.

\textbf{Efficiency measurement.}
We measure end-to-end lesson preparation time using in-system logs collected during teacher-led sessions. The evaluation focuses on the total time required to produce a complete 45-minute lesson compliant with Official Dispatch~5512, including all teacher review and revision stages.

\textbf{User surveys.}
We conduct structured post-task surveys with 18 teachers and 214 students (grades~6--12). Surveys use five-point Likert scales to assess curriculum alignment, lesson clarity, instructional material quality, ease of use, and overall satisfaction. We report means with standard deviations and apply Wilcoxon signed-rank tests to compare teacher satisfaction scores across system variants in the ablation study.

\textbf{Scope of evaluation.}
We explicitly note that the current evaluation does not include a controlled learning outcome study. Consequently, we do not make causal claims about improvements in student learning; survey responses reflect perceived utility and satisfaction rather than measured learning gains. We view a rigorous pre/post study as important future work.

\subsection{Training Dynamics and Subject-wise Performance}
VietEduQwen exhibits stable and consistent improvements throughout four epochs of fine-tuning. As shown in Table~\ref{tab:epoch_accuracy}, overall examination accuracy increases from 80.92\% for the base Qwen3-8B model to 87.02\% after Epoch~4, corresponding to a total gain of +6.10\%. In comparison with representative baselines, VietEduQwen clearly outperforms open-weight models such as Gemma-2-9B and achieves accuracy comparable to Gemini~3~Flash.

\begin{table*}[h]
\caption{Model accuracy across training epochs and baseline comparisons.}
\label{tab:epoch_accuracy}
\centering
\begin{tabular}{lcc}
\hline
Model / Epoch & Accuracy (\%) & Accuracy vs. Base Qwen3-8B \\
\hline
Base Qwen3-8B (Epoch 0) & 80.92 & -- \\
VietEduQwen (Epoch 1) & 83.47 & +2.55 \\
VietEduQwen (Epoch 2) & 85.61 & +4.69 \\
VietEduQwen (Epoch 3) & 86.88 & +5.96 \\
\textbf{VietEduQwen (Epoch 4)} & \textbf{87.02} & \textbf{+6.10} \\
\hline
Gemma-2-9B & 78.73 & $-2.19$ \\
Gemini-2.5-Flash & 85.44 & +4.52 \\
Gemini-2.5-Pro & 88.67 & +7.75 \\
Gemini-3-Flash & 87.65 & +6.73 \\
Gemini-3-Pro & 92.54 & +11.62 \\
\hline
\end{tabular}
\end{table*}

\begin{figure*}[h]
    \centering
    \includegraphics[width=\textwidth]{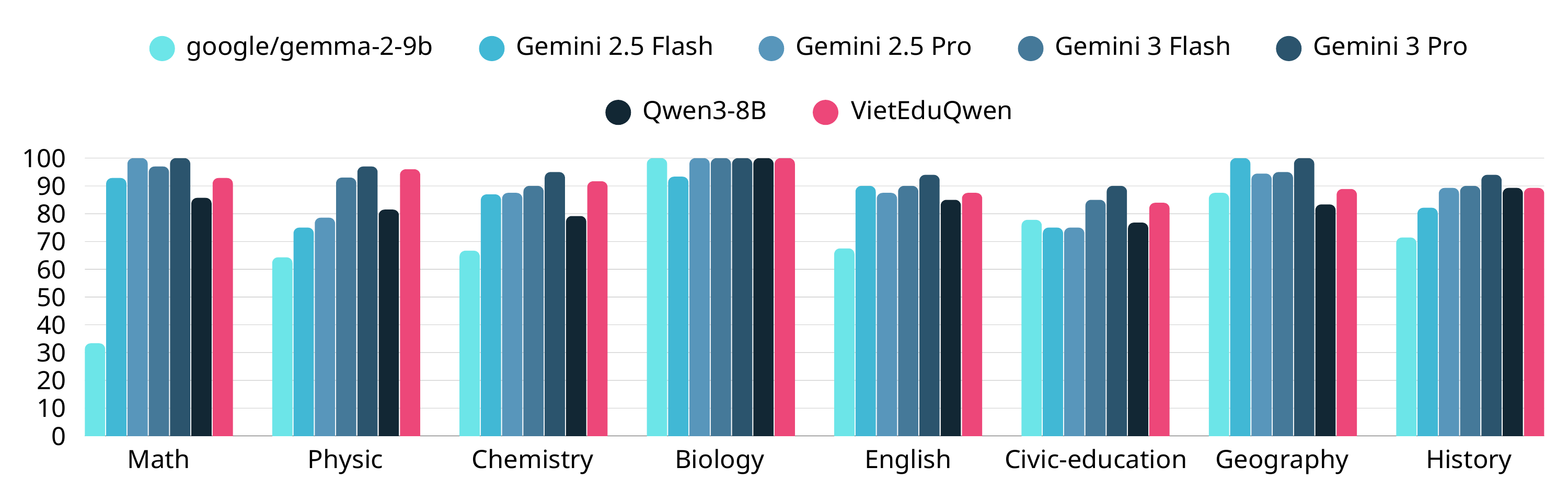}
    \caption{Subject-wise examination accuracy (\%) across all eight subjects.}
    \label{fig:subject_accuracy_split}
\end{figure*}

Fig.~\ref{fig:subject_accuracy_split} presents a detailed subject-wise analysis across eight Vietnamese national examination subjects. VietEduQwen consistently improves over the base Qwen3-8B model across all subjects. The largest gains are observed in STEM disciplines, particularly Mathematics, Physics, and Chemistry, where VietEduQwen substantially narrows the performance gap with Gemini-2.5 and Gemini-3 variants. In Biology, the model reaches near-saturation performance, matching the strongest commercial baselines. For language and social science subjects, including English, Geography, and History, VietEduQwen maintains stable accuracy within a narrow margin of top-performing models. 

To characterize remaining failure modes, we manually categorize a random sample of 200 errors from VietEduQwen's incorrect predictions. Approximately 41\% involve multi-step numerical computation where intermediate rounding or sign errors propagate to incorrect final answers. A further 33\% involve questions requiring integration of information across two or more textbook concepts, where the model correctly identifies each concept in isolation but fails to synthesize them. The remaining 26\% are distributed across reading comprehension errors (Literature and History), unit conversion mistakes (Physics and Chemistry), and ambiguous distractors. These findings suggest that targeted improvements in multi-step reasoning and cross-concept integration represent the most impactful directions for future model development.

\subsection{Instructional Efficiency and User Satisfaction}

Table~\ref{tab:prep_time} reports the average preparation time required to produce a complete 45-minute lesson. ConnectED achieves a substantial efficiency gain, reducing lesson planning time from 3--4 hours under manual preparation to approximately 15--20 minutes, corresponding to a 9--12× speedup. Compared to generic LLM-assisted workflows, ConnectED further minimizes teacher workload by structuring generation around curriculum-aligned templates and reusable components. Importantly, most of the remaining time is spent on reviewing and approving generated outputs rather than authoring content from scratch. This efficiency improvement is achieved without any continuous monitoring or tracking of students during classroom use, resulting in effectively zero student surveillance overhead.

\begin{table*}[h]
\caption{Average preparation time for a 45-minute lesson.}
\label{tab:prep_time}
\centering
\begin{tabular}{lc}
\hline
Method & Preparation Time \\
\hline
Manual preparation & 3--4 hours \\
Generic LLM-assisted workflow & 1--1.5 hours \\
\textbf{ConnectED} & \textbf{15--20 minutes} \\
\hline
\end{tabular}
\end{table*}

\begin{figure*}[h]
    \centering
    \includegraphics[width=\textwidth]{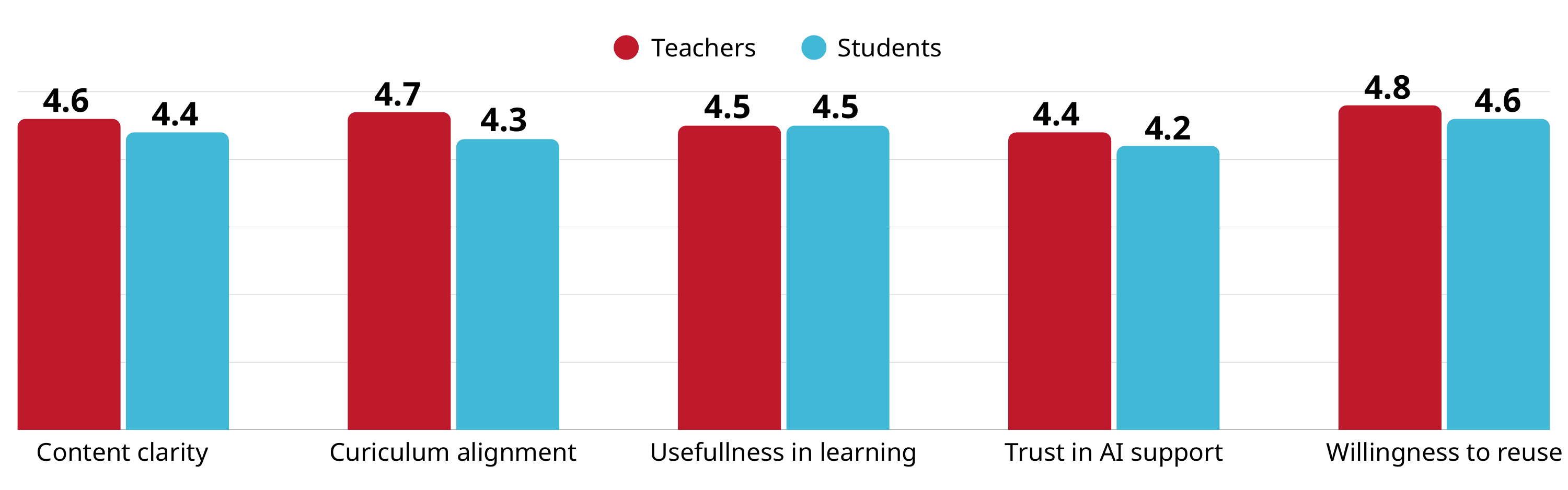}
    \caption{Teacher and student satisfaction scores (mean/5)}
    \label{fig:satisfaction}
\end{figure*}

In addition, user experience evaluations indicate strong acceptance of the system by both teachers and students. Participants rated their experience across five dimensions, with the highest scores observed for curriculum alignment and clarity of instructional materials (Fig.~\ref{fig:satisfaction}). Notably, teachers reported a high level of trust in AI-generated outputs, with 94\% indicating their willingness to reuse ConnectED in future lesson preparation. These findings suggest that ConnectED not only reduces instructional preparation time but also delivers outputs that align well with pedagogical expectations and classroom needs.

\subsection{Ablation Study}

Table~\ref{tab:ablation_model} reports the contribution of individual training components. Removing the DPO objective reduces accuracy by 1.11~percentage points (from 87.02\% to 85.91\%) and leads to qualitatively degraded pedagogical tone, as assessed by two independent annotators rating a sample of 50 responses on a three-level scale (high~/ medium~/ low). The base Qwen3-8B model without any fine-tuning performs substantially worse in both accuracy and pedagogical quality, confirming the importance of domain-specific training.

\begin{table*}[h]
\caption{Model-level ablation on examination accuracy and annotated
  pedagogical quality (two independent annotators, $n{=}50$ sampled
  responses).}
\label{tab:ablation_model}
\centering
\begin{tabular}{lcc}
\hline
Configuration & Accuracy (\%) & Pedagogical Quality \\
\hline
Full VietEduQwen (DPO + SFT)      & \textbf{87.02} & \textbf{High} \\
VietEduQwen without DPO (SFT only) & 85.91          & Medium        \\
Base Qwen3-8B                      & 80.92          & Very low      \\
\hline
\end{tabular}
\end{table*}

Table~\ref{tab:ablation_system} reports the effect of key system components on preparation time and teacher satisfaction. Removing the teacher-in-the-loop review stage reduces preparation time marginally but causes a statistically significant drop in teacher satisfaction ($p{<}0.05$, Wilcoxon signed-rank test~\cite{woolson2007wilcoxon}), indicating that teachers value control over AI outputs even at the cost of additional time. Removing the curriculum checker similarly reduces satisfaction, confirming that automated compliance verification provides meaningful assurance. Replacing the full ADDIE orchestration layer with unstructured prompting of the same VietEduQwen model (the \textit{Generic LLM workflow} row) substantially increases preparation time and achieves the lowest satisfaction score---directly demonstrating that the ADDIE-based orchestration contributes independent value beyond model capability alone. This finding provides empirical support for the framework selection rationale presented in Sect.~\ref{sec:relatedwork}.

\begin{table*}[h]
\caption{System-level ablation on preparation time and teacher satisfaction
  (mean $\pm$ SD; $n{=}18$; $p$-values from Wilcoxon signed-rank test
  vs.\ full pipeline).}
\label{tab:ablation_system}
\centering
\begin{tabular}{lccc}
\hline
System Variant
  & Prep.\ Time (min)
  & Teacher Satisfaction
  & $p$-value \\
\hline
ConnectED (full pipeline)              & \textbf{30--45} & $\mathbf{4.7\pm0.4}$ & --        \\
\quad w/o teacher-in-the-loop          & 25--35          & $3.9\pm0.6$          & $<0.05$   \\
\quad w/o curriculum checker           & 28--40          & $3.8\pm0.7$          & $<0.05$   \\
Generic LLM workflow      & 90--120         & $3.6\pm0.8$          & $<0.01$   \\
\hline
\end{tabular}
\end{table*}

\section{Discussion and Limitations}
\label{sec:discussion}

ConnectED demonstrates strong potential for real-world deployment in Vietnamese secondary education by addressing the gap between the generative capability of large language models and the structured, regulation-driven nature of instructional practice. By aligning lesson generation with Official Dispatch~5512 and embedding teacher validation checkpoints throughout the ADDIE workflow, the system moves beyond one-shot content generation toward a structured instructional support paradigm. In practice, this enables teachers to substantially reduce lesson preparation time while maintaining control over pedagogical decisions. The integration of reusable lesson artifacts and a web-based interface further enhances scalability, allowing educators to standardize lesson quality and iteratively refine instructional materials over time.

Beyond supporting teachers, ConnectED extends its impact to the student learning process through the Playground environment, which facilitates interactive engagement, guided feedback, and continuous collection of learning signals. This design introduces a lightweight yet practical form of learning analytics, enabling teachers to identify learning gaps and adjust instruction accordingly without requiring additional infrastructure. As a result, the system contributes to a more connected instructional loop, where lesson design, delivery, and refinement are closely integrated. Such a paradigm is particularly valuable in contexts where formal assessment systems are limited but continuous pedagogical improvement remains essential.

From a broader perspective, ConnectED aligns with human-centered AI principles that emphasize augmentation rather than replacement of human expertise. By preserving teacher agency through explicit review mechanisms, the system mitigates risks associated with over-reliance on automated generation. At the same time, the use of pedagogically aligned LLM outputs has the potential to improve instructional consistency across classrooms, particularly in under-resourced settings where access to high-quality teaching materials may be uneven. If deployed at scale, systems like ConnectED could contribute to reducing disparities in educational quality and supporting data-informed teaching practices, although such impact depends on responsible deployment and adequate teacher training.

Despite these strengths, several limitations should be acknowledged. First, the current evaluation focuses on examination accuracy, system efficiency, and user satisfaction, which provide useful but indirect indicators of effectiveness. These metrics do not establish causal improvements in student learning outcomes, and therefore the educational impact of ConnectED remains to be validated through controlled pre/post experimental studies. Second, although the training data is carefully curated from textbooks, examination materials, and filtered external sources, the possibility of data leakage between training and evaluation benchmarks cannot be entirely excluded. Future evaluations should adopt temporally disjoint datasets and more rigorous benchmarking protocols to ensure robustness.

In addition, while the ADDIE-based orchestration improves structure and reliability, it may also introduce a degree of rigidity in lesson generation. Experienced teachers may prefer more flexible workflows that allow adaptation beyond predefined instructional phases, suggesting a need to balance structured guidance with user-driven customization. Similarly, the multi-agent framework for STEM visualization prioritizes technical correctness and pedagogical alignment but does not explicitly evaluate cognitive load or learner comprehension. As a result, automatically generated visualizations may vary in effectiveness depending on the learner’s background and context, highlighting the need for further empirical validation in real classroom settings.

Finally, practical deployment challenges remain, including computational cost, internet accessibility, and varying levels of teacher familiarity with AI tools. Although the system is designed for efficiency, these factors may limit adoption in rural or under-resourced environments. Addressing such constraints will be essential for ensuring equitable access and large-scale applicability.

\section{Conclusions and Outlook}
\label{sec:conclusion}

This paper presents ConnectED, a human-centered AI system for curriculum-aligned lesson planning and interactive student learning in Vietnamese secondary education. The system makes two primary contributions including (1) VietEduQwen, a Vietnamese educational LLM trained via SFT and DPO for academic accuracy, pedagogical appropriateness, and student-safe behavior; and (2) an ADDIE-based orchestration framework that operationalizes the ADDIE instructional design model for LLM-driven lesson generation, enforcing Official Dispatch~5512 compliance at each phase boundary, embedding teacher validation gates at every phase transition, and closing a feedback loop from student learning signals back to lesson refinement. Evaluation on the 2025 Vietnamese National High School Examination demonstrates a 6.10\% accuracy gain over the Qwen3-8B base model. Teacher and student surveys report high satisfaction with curriculum alignment and lesson quality, and the full pipeline reduces lesson preparation time by approximately 5--8$\times$ compared to manual workflows. Ablation analyses confirm the independent contributions of the DPO training objective and the ADDIE orchestration layer, with the system-level ablation showing that replacing ADDIE-structured orchestration with unstructured prompting substantially degrades both preparation efficiency and teacher trust. We acknowledge two important limitations. First, the current evaluation relies on examination accuracy and user satisfaction surveys; a controlled study measuring actual student learning outcomes remains future work. Second, data leakage between training materials and the evaluation benchmark cannot be fully excluded, and future evaluations should use temporally disjoint test sets. For future work, we aim to expand ConnectED to rural and under-resourced school settings, explore bilingual instruction, and develop rigorous formative assessment integration that enables evidence-based curriculum refinement. We also plan to conduct a longitudinal classroom study to provide causal evidence of ConnectED's impact on teacher workload and student learning.

\section*{Acknowledgements}

The ConnectED system won first prize in the \selectlanguage{vietnamese} ``A.I Thực Chiến 2025'', \selectlanguage{english} Vietnam’s first national AI competition organized by VTV, NDC, NDA and Techcombank (https://thucchien.ai)

\section*{Declarations}

\begin{itemize}
\item Funding

The authors did not receive support from any organization for the submitted work.

\item Competing interests

The authors declare that they have no conflicts of interest.

\item Ethics approval and consent to participate

Ethical approval was not required for this study in accordance with institutional guidelines, as it involved anonymous and voluntary surveys without collecting sensitive personal data. 

All procedures were conducted in accordance with relevant guidelines and regulations. 
Informed consent was obtained from all participants. 
For participants under the age of 18, written informed consent was obtained from their parents or legal guardians. 
All data were collected anonymously, and no personally identifiable information was recorded.

\item Consent for publication

Not applicable.

\item Data availability

All evaluation settings and prompt templates are available at~\url{https://anonymous.4open.science/r/connected}

\item Contribution of authors

All authors contributed to the conception and design of the study. Material preparation, data collection, and analysis were performed by Thang Doan Viet, Anh Nguyen Hoang, Tinh Luong Son, Anh Hoang Thi Ngoc, and Huyen Giang Thi Thu. The first draft of the manuscript was written by Thang Doan Viet and Tai Le Quy; and all authors commented on previous versions of the manuscript. All authors read and approved the final manuscript.

\end{itemize}

\nocite{*}

\bibliography{sn-bibliography}

\end{document}